\documentclass[pdflatex,sn-mathphys-num]{sn-jnl}%

\usepackage{graphicx}%
\usepackage{multirow}%
\usepackage{amsmath,amssymb,amsfonts}%
\usepackage{amsthm}%
\usepackage{mathrsfs}%
\usepackage[title]{appendix}%
\usepackage{xcolor}%
\usepackage{textcomp}%
\usepackage{manyfoot}%
\usepackage{booktabs}%
\usepackage{algorithm}%
\usepackage{algorithmicx}%
\usepackage{algpseudocode}%
\usepackage{listings}%
\usepackage{url}%

\theoremstyle{thmstyleone}%

\theoremstyle{thmstyletwo}%

\theoremstyle{thmstylethree}%

\begin{document}

\title[CoolPath Tool: A Thermal Comfort Path Planning Tool for Urban Mobility]{CoolPath Tool: A Thermal Comfort Path Planning Tool for Urban Mobility}

\author*[1]{\fnm{Aditya} \sur{Patel}}\email{dev.niyogi@jsg.utexas.edu}
\author[1]{\fnm{Naveen} \sur{Sudharsan}}
\author[1]{\fnm{Trevor} \sur{Brooks}}
\author[1]{\fnm{Harsh} \sur{Kamath}}
\author[2]{\fnm{Dru} \sur{Crawley}}
\author[3]{\fnm{Marc} \sur{Coudert}}
\author[3]{\fnm{Zach} \sur{Baumer}}
\author[1]{\fnm{Dev} \sur{Niyogi}}

\affil*[1]{\orgdiv{}, \orgname{The University of Texas at Austin}, \orgaddress{\state{Texas}, \country{USA}}}
\affil[2]{\orgdiv{}, \orgname{Bentley Systems}, \orgaddress{\state{Washington DC}, \country{USA}}}
\affil[3]{\orgdiv{}, \orgname{City of Austin}, \orgaddress{\state{Texas}, \country{USA}}}

\abstract{Extreme heat poses a growing challenge for active transportation in cities where conventional weather reporting (e.g.\ limited air temperature measurement for the whole city) fails to capture the large microclimate variations that pedestrians and cyclists experience. We present a novel walking and biking route planning tool (``CoolPath Tool'') that selects paths based on thermal comfort using the Universal Thermal Climate Index (UTCI) rather than just distance or travel time. This system combines high-resolution thermal modeling with real-time route mapping. We generate city-scale UTCI maps using GPU version of Solar and LongWave Environmental Irradiance Geometry (SOLWEIG) model, to account for urban features (buildings, trees,) and weather conditions. The urban features are pre-mapped using satellite data products and the routes are the roadways. For any given origin and destination, our tool calculates the average UTCI along each possible route and recommends the ``coolest'' route, i.e.\ the path with the lowest heat stress (often the most shaded or otherwise thermally comfortable), while still being reasonably direct. We demonstrate this in a case study for Austin, Texas. The approach identifies routes that significantly reduce pedestrians' heat exposure (often recommending routes with a much larger proportion of shade). Such thermally-informed route planning has important public well-being, and economic implications: by helping people avoid dangerous heat hotspots and sun-exposed areas, it can reduce the risk of heat-related illness and make walking or biking a safer choice even on hot days. This tool is part of the Austin Digital Twin efforts developed as part the UT-City CoLab needs. The work while demonstrated for Austin, TX is scalable, and transferrable to other cities globally.}

\maketitle

\section{Introduction}\label{sec1}

Extreme heat is the leading weather-related killer in the United States, causing 177 mortalities in 2024 alone \cite{nws2024}. It has been well documented that extreme heat is exacerbated within urban areas around the world due to the urban heat island effect \cite{mills2022}. The abundance of man-made surfaces such as concrete and asphalt, lack of shading, differences in urban form, and increased concentrations of anthropogenic heat waste all collectively enhance urban heat, and thereby accelerate its negative effects \cite{stone2010}. While extreme heat has a myriad of negative impacts, including on infrastructure \cite{guddanti2025}, the economy \cite{yuan2024}, and natural wildlife \cite{brans2017}, one of its most alarming effects is on human health. Extreme heat's effects on health have been well documented. These include the exacerbation of existing symptoms and underlying diseases (such as diabetes), increased risk of hospitalization due to heat related illnesses, as well as an increased risk of death \cite{basu2009}. While the greatest risk is for individuals with preexisting conditions, the elderly, young children, even healthy, fit, and young individuals can face negative consequences to prolonged heat exposure and have a broader negative impact on the economy \cite{miner2017}.

These health risks are not just a function of individual susceptibility but also are dependent on exposure levels. A common pathway for exposure to extreme heat in cities is link into urban mobility. Those that commute via walking, biking, or public transportation often experience significant exposure to extreme heat \cite{seong2025}. Importantly, the routes which these commuters choose to take may have a drastic effect on their exposure level. Intuitively, people understand that things such as increased wind speed, access to shading, and proximity to green and blue space will likely decrease their exposure and likelihood of experiencing a negative health outcome. A recent study in Miami supports this intuition, finding that only a few streets and bus stops caused most of the heat exposure within their city \cite{liu2025}. This means that the if commuters are uneducated in the location of these ``hot routes'' they may unnecessarily be exposing themselves to intensely hot microclimates, when cooler paths may be available to them. This is likely exacerbated as traditional route planning apps (Google and Apple Maps) prioritize efficiency in commute time and currently do not consider potential heat outcomes.

This article introduces a tool in the form of an interactive thermal comfort walking path planner. The planner operationalizes the idea that different paths will have different heat exposure levels for an individual non-motorized commuter by presenting and ranking potential routes based on UTCI (universal thermal climate index) as a heat index. The UTCI metric is used as it is more akin to what a person feels as they are navigating the thermal environment. It is derived from a variety of metrics such as humidity, wind speed, and mean radiant temperature. The first approach was to create a high resolution UTCI data product (2-meter grid resolution). This dataset, called, ThermalScape \cite{kamath2026} is used as it captures the cooling effects of individual trees, shading and reflection due to buildings and measures it on a scale that is directly relevant to exposure outcomes. The route planner leverages UTCI and calculates route options that are optimum for non-motorized commuters to take to limit heat exposure.

\section{Methodology}\label{sec2}

\subsection{Overview}\label{sec2a}

The coolest route planner integrates an urban climate modeling framework Solar and LongWave Environmental Irradiance Geometry (SOLWEIG-GPU) with a mapping application. There are two main components: (1) City-wide UTCI computation with SOLWEIG-GPU to capture the spatiotemporal distribution of heat stress index UTCI, and (2) Route analysis to compute and compare the thermal exposure of candidate paths queried using Google maps. The system is implemented with a Python-based backend (Flask API) that handles data processing and a frontend (React/CesiumJS) for interactive 3D visualization. Below we detail each part of the process.

\subsection{High-Resolution UTCI Mapping}\label{sec2b}

We used the SOLWEIG-GPU model to generate fine-scale UTCI maps for Austin, Texas. SOLWEIG is an established urban microclimate model that calculates key variables like Sky View Factor (SVF) (a measure of how much of the sky is visible, inversely related to shading), Mean Radiant Temperature (MRT) (combined effect of all radiant heat fluxes), and ultimately UTCI \cite{jendritzky2012}. The GPU-accelerated version of SOLWEIG enables running these calculations at a 2-m resolution across city-scale domains. The modeling workflow is defined as follows: first, the city is divided into tiles, and the input data layers are prepared, including (i) a Building Digital Surface Model (DSM) that encodes the height of buildings and terrain, (ii) a Digital Elevation Model (DEM) for ground elevation (terrain without buildings), (iii) a Tree canopy model (DSM of vegetation height), and (iv) land cover information (surface material types). Key meteorological variables including air temperature, relative humidity, wind speed (10 m wind components), atmospheric pressure, and solar radiation (shortwave and longwave) are included as inputs. Using these inputs, SOLWEIG-GPU computes, for each grid cell in the city, the relevant radiative fluxes (accounting for shadows of buildings/trees each hour, reflection, emission, etc.) and the resulting UTCI value at pedestrian level ($\sim$2 m height) \cite{kamath2023}. The computationally intensive steps, namely the calculation of Skyview Factor (SVF) are offloaded to the GPU, resulting in a drastic speed up in the generation of rasters. The final output is a high resolution UTCI raster for a given time. These maps reveal the spatial variation in heat stress across Austin -- for example, at 4 PM, downtown streets with tall buildings might show UTCI $\sim$35$^\circ$C (in the strong heat stress range) in the shade of skyscrapers, whereas an open parking lot might exceed 42$^\circ$C (approaching very strong heat stress), even if the ambient air temperature is the same in both locations. These differences align with intuition: areas with shade have consistently lower UTCI, and as a result show lower heat stress. Such differences align with known patterns: areas without shade have the highest UTCI, while areas under tree shade show significantly lower heat stress.

\begin{figure}[h]
\centering
\includegraphics[width=0.9\textwidth]{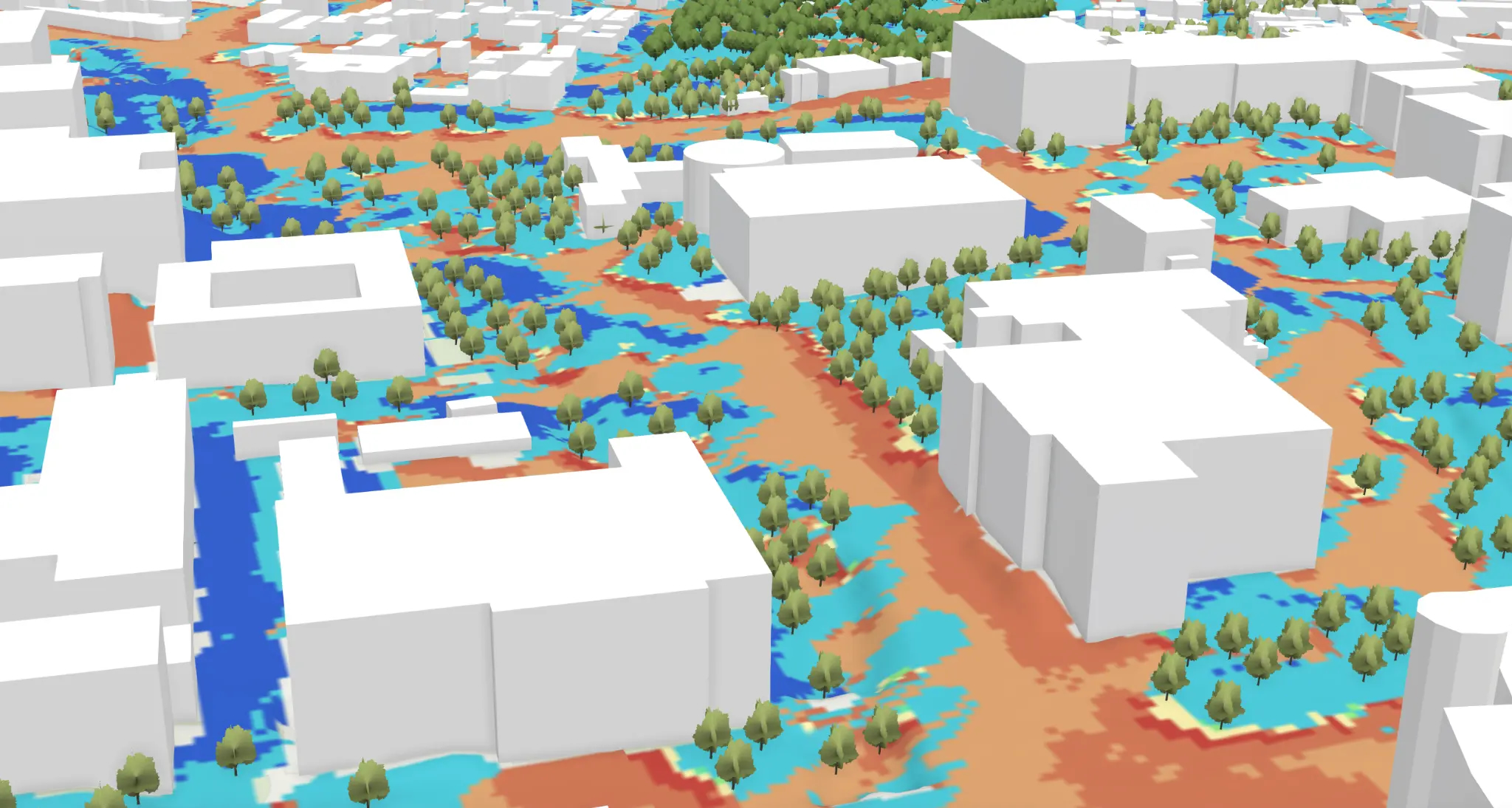}
\caption{3D Representation of UTCI Map for part of a neighborhood in Austin, Texas. The warmer brown colors indicate higher heat stress while the blue and teal indicate shades and relatively cooler locations.}\label{fig1}
\end{figure}

\subsection{Route Extraction}\label{sec2c}

To calculate viable paths for a given route, we utilize the Google Maps Directions API. In this the user enters an origin and a destination, the backend sends a request to Google's service and retrieves multiple parts (typically 2--3 routes, sometimes more) that are considered reasonable by Google Maps.

The JSON response is parsed to extract each route's geometry (as an encoded polyline of latitude/longitude points) and metadata like distance and travel time. The polyline is then decoded into a sequence of lat--lon coordinates for the path. However, the set of points received from the API are sparse (containing mostly only turning points). For accurate analysis, UTCI values must be sampled along the entire route, therefore we densify the route geometry by interpolating points at a regular interval (we chose every $\sim$4 m). This ensures that all the small-scale variations in the underlying UTCI data are captured effectively. The interpolation is done in a geodesic manner: we compute segment lengths and insert intermediate points such that no gap between sample points exceeds the target distance. The result is a list of hundreds of points per route, spaced roughly uniformly.

\subsection{Thermal Exposure Calculation}\label{sec2d}

Each route's points are converted to the coordinate system of the raster and overlaid on the UTCI grid. This yields a sequence of UTCI values for all the points along the route. We then compute summary statistics including the average UTCI for that route, which serves as our indicator of overall heat stress on that path. We also record the minimum and maximum UTCI encountered (showing the coolest and hottest spots on the route), and we calculate a ``shade percentage'' metric: the proportion of the route's length that is in relatively cooler, shaded conditions. To estimate this, we take a threshold defined as the lower ``X\%'' of UTCI values (i.e.\ the coolest end of the spectrum) and count what fraction of a route's points fall below that threshold. In our analysis, we use the 90th-percentile of the UTCI range as a cutoff. This proxy value is useful in calculating the relative difference between given routes in terms of benefit from shade or other cooling factors. The route computations are implemented in the Flask API which returns for each route: the coordinates, mean/min/max UTCI, normalized thermal score, shade percentage, distance, and duration.

\subsection{Visualization and User Interface}\label{sec2e}

The frontend (built with React and CesiumJS) is designed as an intuitive way for users to interact with the planner. Users enter their origin and destination (address or place name) into a form. In our example we used the University of Texas at Austin Tower, to a location in downtown ``1100 Congress Avenue''. The app geocodes these to coordinates using Google's Geocoding API and then requests the backend for route processing. Once results are received (typically in a few seconds), the app displays the routes on a 3D map. We use CesiumJS to render a realistic model of the city: terrain and photorealistic 3D building tiles (via Google's 3D Tiles API) are loaded so the user sees the actual urban landscape in 3D. The routes are overlaid as color-coded polylines: We assign a color gradient from red (hottest route) to blue (coolest route), with intermediate routes in green, etc., corresponding to their relative mean UTCI. Each route line can be clicked to show a popup with its details. A legend is provided to interpret colors (e.g.\ red = ``Warm'', blue = ``Coolest''). Additionally, the map shows markers for the start and end points and automatically zooms to the route bounds. Users can thus see, for example, one route going through a shaded park (drawn in blue) versus another along a sun-exposed main road (drawn in red). This visual comparison, backed by the quantitative UTCI analysis, makes it easy to understand which route will likely be more comfortable. The entire application is designed to be interactive and near real-time. This is done by utilizing the hourly forecast from the previous day for the meteorological inputs in the SOLWEIG-GPU model.

\section{Prototype}\label{sec3}

\begin{figure}[h]
\centering
\includegraphics[width=0.9\textwidth]{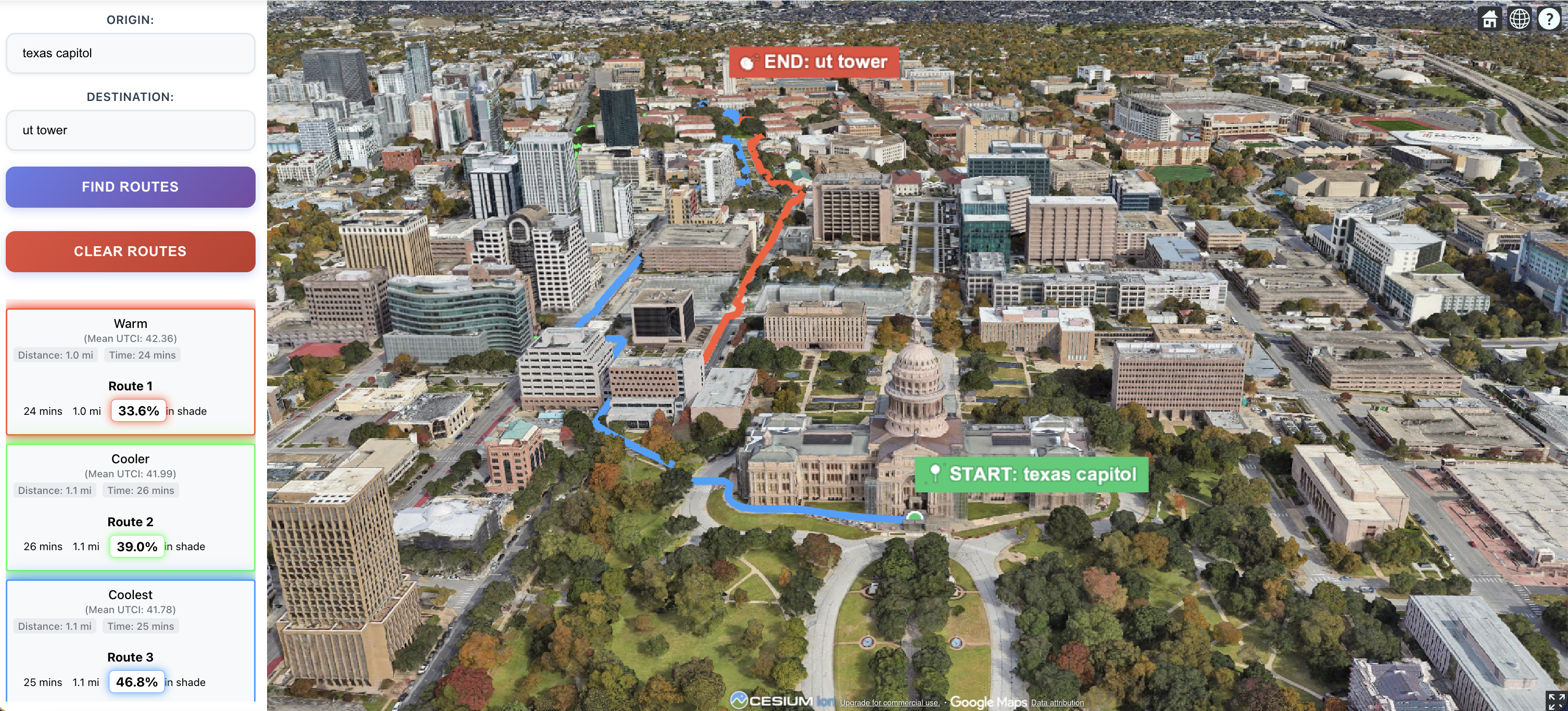}
\caption{Snapshot of the Interactive Path Planning Tool showing the CoolPath Tool example for Austin, Texas.}\label{fig2}
\end{figure}

\subsection{Example Case Study -- Thermally Optimal Paths}\label{sec3a}

\begin{figure}[h]
\centering
\includegraphics[width=0.3\textwidth]{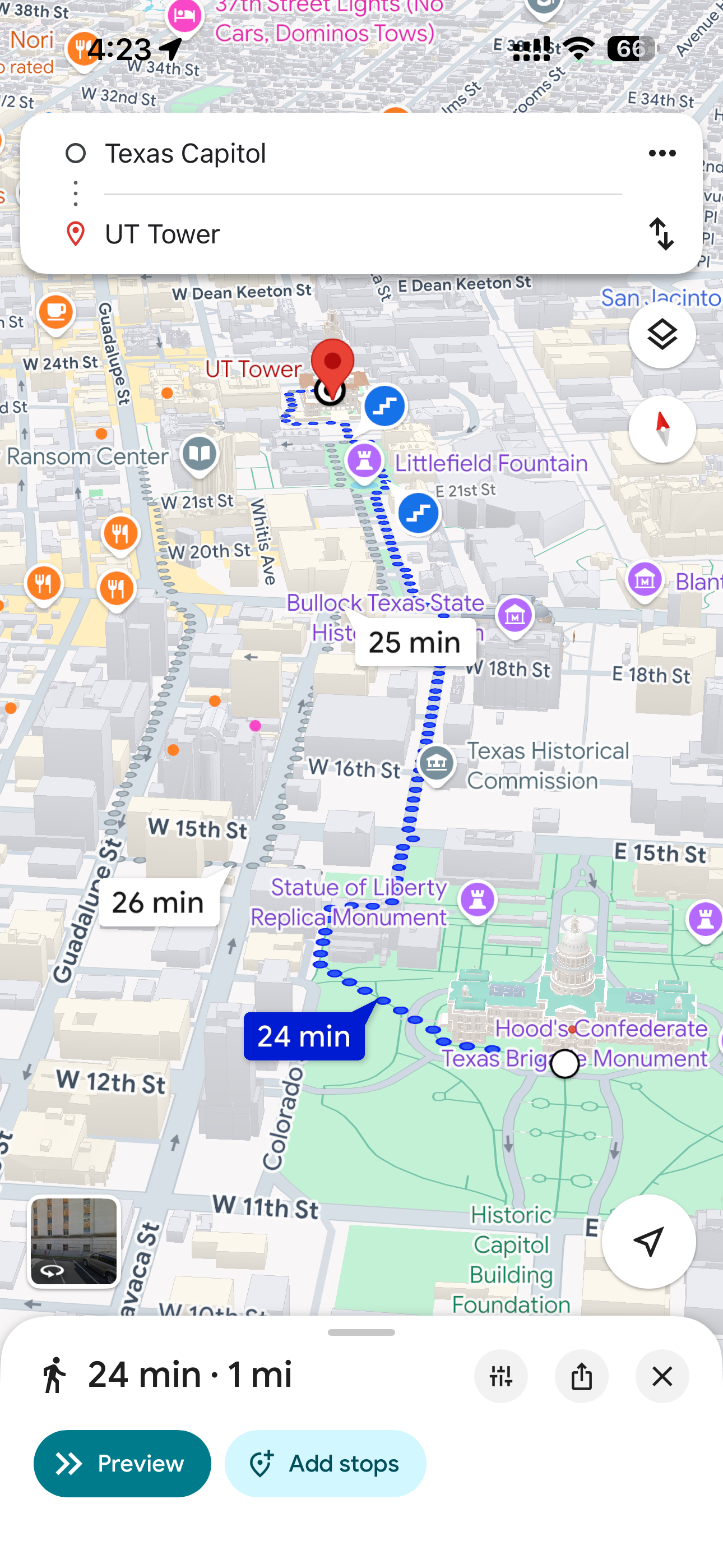}
\caption{Walking route suggested by Google maps for the same origin-destination pair.}\label{fig3}
\end{figure}

To demonstrate the planner's capabilities, we consider an afternoon scenario in Austin (around 4 PM on a clear summer day). The origin selected is the Texas Capitol and the destination is selected as the University of Texas at Austin Tower, at an approximate distance of 1 mile. The planner calculated three alternate walking routes. Figure~\ref{fig2} illustrates these three routes on the map, color-coded by heat stress level (red = most heat exposure, blue = least heat exposure). While the difference in the length of the three routes is negligible, Route 1 (red) had about 33.6\% of the route classified as shaded, whereas Route 3 (blue) had 46\% of the route classified as shaded. Predictably, the warmer routes follow along wider, larger streets with limited tree cover and shorter buildings, whereas the cooler route follows inner streets which benefit from shade from buildings and tree cover. While the lengths of the routes are similar, when intuitively selecting a route, most individuals would avoid the `coolest path' as found by our tool as it appears more indirect, and less straightforward, highlighting that the most thermally optimal routing is unlikely to be discovered by visual inspection alone. Figure~\ref{fig3} shows the same path as outlined by the Google Maps application. Notably, the recommended path corresponds to the highest heat stress among all the routes.

\subsection{Scalability and Timing}\label{sec3b}

Using precomputed UTCI rasters for the city, the route analysis is very efficient. The most computationally intensive component (the SOLWEIG-GPU simulation) is done a priori. It can take up to 2--3 hr to generate a full day of hourly UTCI maps for the entire city at high resolution, and our efforts are underway to optimize this further using the GPU approach. However, once these maps have been generated, each route request involves simple computations which take a few seconds on a typical server. In our demonstration, the end-to-end response time after clicking ``Find Routes'' was usually around 2--5 seconds for computing three alternatives. This means the approach can scale to real-time deployment across cities, updating thermal maps once per hour, having precomputed 24 hourly maps on the previous day using forecast data.

\subsection{Generality}\label{sec3c}

While our focus was Austin, Texas, the approach is generalizable. The pipeline we have created can be applied to any city for which the following data is available: A digital surface model (from LiDAR or UT-GLOBUS, \cite{kamath2023}, or similar), meteorological inputs (e.g.\ MERRA, UMEP, WRF), and a routing engine (Google or OpenStreetMap-based services).

\section{Discussion}\label{sec4}

The development of an interactive thermal comfort walking path planning tool has several important implications for both public well-being and urban planning. Most directly, it empowers individuals to make informed choices on their commutes to lower overall heat exposure and limit negative health outcomes from extreme urban heat. By showing individuals cooler and shadier routes, the planner can reduce cumulative heat exposure, ensuring that an individual's thermal load does not reach the point in which a negative health outcome is likely. Providing options for people to stay out of the worst heat can be lifesaving at best, informative at worst. In this way, the interactive mapper can be seen as akin to a preventative health measure that people can use to increase their overall health in the summer. This tool could also be likely used in the winter to with the reverse intention in mind--that is, finding the warmest paths to protect from extreme cold and more sun exposure.

The tool could also encourage non-motorized transportation during hotter weather. This would likely have positive effects in terms of people's overall personal fitness and the environment in the form of lower vehicle emissions. One of the paradoxes of extreme heat is that individuals combat it through the use of air conditioning, which in turn exacerbates the heat island \cite{tremeac2012}. If people find that walking paths available to them were relatively cool, even under hot conditions, it is likely to increase pedestrian mobility rather than staying inside or traversing by motorized vehicle. This feature was recently highlighted to the study team from the City of Austin's Pedestrian Advisory Council \cite{austinpac2025}. From a planning perspective, there is evidence that pathways and neighborhoods are overestimated in terms of their walkability when heat is not accounted for. With one study describing how perceived walking distance is negatively affected as heat stress increases \cite{basu2024}. Providing information on shadier and cooler routes can restore the accessibility of certain neighborhoods and important areas within a city that may be ``locked'' during hot weather, and thus help local economy as well.

Continuing to access the tool's efficacy within a planning framework, it can be used as a diagnostic tool for major ingress and egress routes around the city. For example, many cities have implemented some economic programs focused on developing economic corridors within the city \cite{cityofaustin2024}. These corridors rely on foot traffic throughout the year from tourists and locals alike to maintain profitability. If people are not walking these corridors within the summer, it could lead to business closures and the failure of a city priority program. This route planner can access the walkability of these economic corridors during high heat events and if they are found to be wanting, it can help incentivize the city to invest in cooling infrastructure for these routes. This may also aid prioritization efforts by showing areas that are systematically devoid of any cooling infrastructure--that is, showing areas where there are no ``cool'' routes due to a lack of tall buildings and greenspace. The tool can help citizens with data to show city planners and decision makers that can back up their lived experiences as commuters as well.

Often, citizens make decisions for commuting based on vague city-wide metrics. For example, looking at the ambient temperature provided to them by their local weather stations. When it comes to what route to take, they may use a map application such as google or apple maps. Heat advisories, while useful are not necessarily realistic for everyone. Some people must commute for work even in extreme heat advisories. Mapping applications, as previously stated, prioritize route efficiency in terms of time rather than health impacts. Even when used in tandem these two tools fail to inform citizens of the local microclimates they will be facing as they walk. The CoolPath Tool offers an alternative providing safer ways to navigate the urban environment even during the hottest periods. The Tool was built to be straight forward, with end users in mind and operates very similarly to a traditional mapping application, only our priorities combine efficiency with heat exposure, lowering the barrier for mobility adaptation at the individual level.

While the tool has much potential, there are also some limitations that need to be addressed. For example, the UTCI model that is powering this tool is built to assume a standard metabolic rate and attire. Meaning that people of different ages, weights, heights, as well as people in different clothing may experience thermal load differently than how the mapper shows. However, while absolute thermal sensation may vary across individuals, the relative differences between alternative routes remain consistent, meaning that routes identified as cooler by the tool should remain comparatively cooler for all users, even if the magnitude of perceived comfort differs. 

Despite these potential limitations, our work shows that incorporating thermal load into route planning is important and possible. The public health aspect of this work cannot be overstated. This has the potential to protect lives, increase economic efficiency, and give end users an overall more comfortable life by arming end users with the tools to adapt to extreme heat in real time. As cities continue to learn to deal with urban heat, and an increasing number and intensity of heat waves, tools like these become even more valuable to allow individuals to protect themselves and make educated decisions. The public planning efficacy of this tool is also important to highlight. Planners can use this to identify where to prioritize heat mitigation infrastructure, especially within the context of important transportation and economic corridors. Future efforts can also be directed with rain and other extremes related mobility \cite{tiwari2025,fakhruddin2022}.

\section{Limitations}\label{sec5}

The current prototype relies on precomputed static UTCI datasets rather than dynamically updated real-time data. Although this is sufficient for demonstration and proof-of-concept purposes, a fully dynamic implementation would require frequent recomputation of high-resolution city-scale UTCI rasters, which incurs substantial computational overhead and storage costs, particularly when scaling to multi-day or multi-city deployments. We are actively working to reduce computational costs through algorithmic optimizations and improved memory management.

\section{Conclusions}\label{sec5}

This work demonstrates a practical way to reduce heat exposure to individuals by leveraging microclimate modeling and route optimization. It emphasizes the importance of shading infrastructure to protect citizens in a warm climate. We showed that within a city, the importance of choice can be the difference between significant levels of thermal load. These choices can and should be made by the proper tools both on a city planning level and at the individual level. By creating a user-friendly tool our work should assist in improving the walkability of neighborhoods and comfortability of residents. For our study domain of Austin, Texas the work helped the Pedestrian Advisory Council to provide recommendations about using such technology in developing sidewalks programs that consider exposure and help build ``mitigating infrastructure alongside sidewalk improvements'' for improving pedestrian and bicycle mobility.

Moving forward, this work will hope to improve upon the tool by becoming more dynamic and expanding the tool to other cities. Especially those in the southwest and the global south where beating the heat goes hand in hand with promoting active and sustainable development.

In empowering commuters to identify thermally optimum routes, we not only hope to reduce instances of negative health outcomes, but also to encourage non-motorized urban commuting even in hot weather, and spark discussion on the importance of thermally beneficial infrastructure along important non-motorized commuter routes.

\backmatter

\section*{Declarations}

\subsection*{Competing interests}
The authors declare that they have no conflict of interest.

\subsection*{Data availability}
The sample UTCI dataset used in this study is publicly available on Zenodo at:  
https://zenodo.org/records/18283037

\subsection*{Code availability}
The SOLWEIG-GPU package used to generate the dataset is publicly available on GitHub at:  
https://github.com/nvnsudharsan/SOLWEIG-GPU
\subsection*{Author Contributions}
A.P. :  overall  design, system implementation, analysis, and manuscript preparation. \\
H.K. and N.S.:  Developing SOLWEIG-GPU framework and  the high-resolution UTCI datasets used in this study. \\
T.B.:  System design, visualization workflow, and assisted with manuscript writing and editing. \\
D.C.: Technical guidance on the CesiumJS-based visualization and 3D mapping components. \\
M.C. and Z.B.:  Domain expertise and provided  input from the City of Austin perspective. \\
D.N.: supervised the project, provided strategic direction, and contributed to the conceptual framework,  interpretation of results, editing the manuscript. \\
All authors reviewed and approved the final manuscript.

\subsection*{Acknowledgements}

This work benefitted in part through National Aeronautics and Space Administration Disasters Program [80NSSC25K7417], National Institutes of Standards and Technology [60NANB24D235], U.S. National Science Foundation (NSF) [AGS 2413827, 2324744, 2435096], UT-City CoLab, UNESCO Chair at UT Austin Jackson School Geosciences, UT-UNESCO International Initiative (U2I2, S. Kumar and R. Bashyam Family gift), Farish Endowment, and Bentley Systems DAF. The simulations were conducted on the Stampede3 supercomputer at the Texas Advanced Computing Center (TACC). We thank the City of Austin Pedestrian Advisory Council for their feedbacks that helped the CoolPath Tool design.

\bibliography{sn-coolpath-bibliography}

@misc{austinpac2025,
  author = {{Austin PAC}},
  year = {2025},
  title = {City of Austin Pedestrian Advisory Council, Recommendation 20251103-006: Supporting the Sidewalks Program with the UT-City CoLab},
  note = {Available from \url{https://services.austintexas.gov/edims/document.cfm?id=462929}}
}

@article{basu2009,
  author = {Basu, R.},
  title = {High ambient temperature and mortality: a review of epidemiologic studies from 2001 to 2008.},
  journal = {Environmental health},
  volume = {8},
  number = {1},
  pages = {40},
  year = {2009}
}

@article{basu2024,
  author = {Basu, R. and Colaninno, N. and Alhassan, A. and Sevtsuk, A.},
  title = {Hot and bothered: Exploring the effect of heat on pedestrian route choice behavior and accessibility.},
  journal = {Cities},
  volume = {155},
  pages = {105435},
  year = {2024}
}

@article{brans2017,
  author = {Brans, K. I. and Jansen, M. and Vanoverbeke, J. and T{\"u}z{\"u}n, N. and Stoks, R. and De Meester, L.},
  title = {The heat is on: Genetic adaptation to urbanization mediated by thermal tolerance and body size.},
  journal = {Global change biology},
  volume = {23},
  number = {12},
  pages = {5218--5227},
  year = {2017}
}

@misc{cityofaustin2024,
  author = {{City of Austin}},
  title = {Souly Austin | AustinTexas.gov},
  year = {2024},
  url = {https://www.austintexas.gov/page/souly-austin}
}

@article{fakhruddin2022,
  author = {Fakhruddin, B. and Kirsch-Wood, J. and Niyogi, D. and Guoqing, L. and Murray, V. and Frolova, N.},
  title = {Harnessing risk-informed data for disaster and climate resilience.},
  journal = {Progress in disaster science},
  volume = {16},
  pages = {100254},
  year = {2022}
}

@article{guddanti2025,
  author = {Guddanti, K. P. and Bharati, A. K. and Nekkalapu, S. and Mcwherter, J. and Morris, S.},
  title = {A comprehensive review: impacts of extreme temperatures due to climate change on power grid infrastructure and operation.},
  journal = {IEEE Access},
  year = {2025}
}

@article{jendritzky2012,
  author = {Jendritzky, G. and De Dear, R. and Havenith, G.},
  title = {UTCI---why another thermal index?},
  journal = {International journal of biometeorology},
  volume = {56},
  number = {3},
  pages = {421--428},
  year = {2012}
}

@article{kamath2023,
  author = {Kamath, H. G. and Martilli, A. and Singh, M. and Brooks, T. and Lanza, K. and Bixler, R. P. and Niyogi, D.},
  title = {Human heat health index (H3I) for holistic assessment of heat hazard and mitigation strategies beyond urban heat islands.},
  journal = {Urban Climate},
  volume = {52},
  pages = {101675},
  year = {2023}
}

@misc{kamath2026,
  author = {Kamath, H. and Sudharsan, N. and Singh, M. and Wallenberg, N. and Lindberg, F. and Niyogi, D.},
  title = {SOLWEIG-GPU: GPU-Accelerated Thermal Comfort Modeling Framework.},
  year = {2026},
  publisher = {Zenodo}
}

@article{liu2025,
  author = {Liu, L. and Li, X. and Pereira, R. H. and Yan, X.},
  title = {Measuring exposure to extreme heat in public transit systems.},
  journal = {Journal of Transport Geography},
  volume = {128},
  pages = {104383},
  year = {2025}
}

@article{mills2022,
  author = {Mills, G. and Stewart, I. D. and Niyogi, D.},
  title = {The origins of modern urban climate science: reflections on `A numerical model of the urban heat island'.},
  journal = {Progress in Physical Geography: Earth and Environment},
  volume = {46},
  number = {4},
  pages = {649--656},
  year = {2022}
}

@article{miner2017,
  author = {Miner, M. J. and Taylor, R. A. and Jones, C. and Phelan, P. E.},
  title = {Efficiency, economics, and the urban heat island.},
  journal = {Environment and Urbanization},
  volume = {29},
  number = {1},
  pages = {183--194},
  year = {2017}
}

@misc{nws2024,
  author = {{National Weather Service and NOAA}},
  title = {Summary of Natural Hazard Statistics for 2024 in the United States.},
  year = {2024},
  url = {https://www.weather.gov/media/hazstat/sum24.pdf}
}

@article{stone2010,
  author = {Stone, B. and Hess, J. J. and Frumkin, H.},
  title = {Urban form and extreme heat events: are sprawling cities more vulnerable to climate change than compact cities?},
  journal = {Environmental health perspectives},
  volume = {118},
  number = {10},
  pages = {1425},
  year = {2010}
}

@article{tiwari2025,
  author = {Tiwari, A. and Cherkauer, K.},
  title = {How much did it rain? A comparison of four datasets for TC Beryl (2012).},
  journal = {Journal of Applied Meteorology and Climatology},
  volume = {64},
  number = {10},
  pages = {1431--1446},
  year = {2025}
}

@article{tremeac2012,
  author = {Tremeac, B. and Bousquet, P. and de Munck, C. and Pigeon, G. and Masson, V. and Marchadier, C. and others and Meunier, F.},
  title = {Influence of air conditioning management on heat island in Paris air street temperatures.},
  journal = {Applied Energy},
  volume = {95},
  pages = {102--110},
  year = {2012}
}

@article{yuan2024,
  author = {Yuan, Y. and Li, X. and Wang, H. and Geng, X. and Gu, J. and Fan, Z. and others and Liao, C.},
  title = {Unraveling the global economic and mortality effects of rising urban heat island intensity.},
  journal = {Sustainable Cities and Society},
  volume = {116},
  pages = {105902},
  year = {2024}
}

@article{seong2025,
  author  = {Seong, K. and Choi, S. J. and Jiao, J.},
  title   = {IoT sensors as a tool for assessing spatiotemporal risk to extreme heat},
  journal = {Journal of Environmental Planning and Management},
  volume  = {68},
  number  = {11},
  pages   = {2621--2643},
  year    = {2025}
}

\end{document}